\documentclass[aps,prd,onecolum,11pt,showkeys]{revtex4-2}

\usepackage{amsmath,amssymb}
\usepackage{graphicx}

\usepackage[colorlinks=true,linkcolor=blue,citecolor=blue,urlcolor=blue]{hyperref}

\newcommand{\IFFC}{Instituto de F\'{\i}sica, Facultad de Ciencias, Igu\'a 4225, esq. Mataojo, 11400 Montevideo, Uruguay.}

\begin{document}

\title{Consistency of the LQG quantization of black holes coupled with scalar matter and a clock}

\author{Rodrigo Eyheralde$^1$}
\author{Rodolfo Gambini$^1$}

\affiliation{$^1$\IFFC}

\email{reyheralde@fisica.edu.uy}


\begin{abstract}
The Dirac quantization of spherically symmetric gravity coupled to a scalar field in Loop Quantum Gravity remains unresolved, mainly because of the difficulty in maintaining a consistent constraint algebra at the quantum level. One possible way to overcome this obstruction is to fix the gauge by coupling the system to a physical clock. However, this approach requires careful control of the consistency of the gauge-fixed theory and factor-ordering ambiguities. Here, we address these issues by analyzing whether the gauge-fixed quantization reproduces the well-known results for the quantization of a black hole in vacuum using the Dirac method. This requires a treatment valid throughout the outer region of the black hole, where the asymptotic approximations considered in previous studies do not hold true.
\end{abstract}
\maketitle
\section{Introduction}
In a previous study \cite{Gambini:2023pox}, the Loop Quantum Gravity (LQG) quantization of spherically symmetric gravity coupled to a scalar field was introduced, using a second scalar field as a physical clock. Thus far, a full Dirac quantization of this system has been impossible owing to difficulties in preserving the constraint algebra in the presence of matter.
The gauge-fixed approach allows one to fix the time variable and describe the system through a \textit{true} Hamiltonian, thereby bypassing the constraint algebra problem while retaining, as we shall see, consistency conditions that help to eliminate ambiguities and identify the correct quantization procedure.\\

Subsequently, in \cite{Gambini:2024suj,Gambini:2023wnc}, the radially asymptotic region of the interacting gravity–scalar system was studied. In that analysis, the underlying discreteness of the theory was lost due to the approximations involved, leading to several ambiguities that are discussed in the following. In particular, the ground state of the Hamiltonian was subject to ambiguities that made it difficult to identify the physical states. This, in turn, suggested a potential consistency between the gauge-fixing procedure and the Hamiltonian constraint. Specifically, we must rigorously show that the energy of the fundamental state of the true Hamiltonian corresponds to the solution of the Hamiltonian constraint. In other words, when the clock energy is negligible, the energy of the fundamental state is also negligible.\\

In the present study, we address the gauge-fixed fully polymerized LQG system, covering the entire exterior region of the black hole (not only the asymptotic region). We begin by focusing on the purely gravitational sector, with no matter content other than the clock field. We resolve the ambiguities in the definition of the true Hamiltonian spectrum, verify the consistency with the Hamiltonian constraint, and lay the groundwork for a numerical analysis of the exact spectrum and quantum effects of matter coupled to gravity.\\

The remainder of this section reviews the LQG construction for spherically symmetric gravity coupled to a clock field. Section \ref{sec:quantum_theory} discusses the Hamiltonian and its spectrum. The key ingredient in this analysis is the determination of the spectrum of the Hamiltonian constraint, whose eigenstates and eigenvalues are computed using a uniform WKB approximation in Sections \ref{sec:Finnite_Diff} and \ref{sec:spectral_analysis}. This result is then used to study the spectrum of the Hamiltonian, with an emphasis on its ground level in Section \ref{sec:Hamiltonian}. In Section \ref{sec:Analytic_approx}, we explore the qualitative features of the spectrum beyond the ground state, and outline the framework for a numerical investigation of the system. Finally, Section \ref{sec:comparison} presents a comparison with Ref. \cite{Gambini:2023wnc}, where the total interacting system is analyzed using more restrictive approximations. Concluding remarks are presented in the final section, and two appendices containing detailed calculations are included to avoid overloading Section \ref{sec:quantum_theory}.

\subsection{Classical system: gravity coupled to scalar matter and a clock}
As a starting point, we consider the total Hamiltonian constraint for spherically symmetric gravity coupled to two scalar fields. For the canonical variables, we follow the established choice in LQG (see \cite{Gambini:2022hxr}). Specifically, we consider the radial $(E^x)$ and tangential $(E^\varphi)$ components of the triad, together with their conjugate momenta $K_x$ and $K_\varphi$ as canonical pairs. For the scalar fields, we use their amplitudes $\phi$ and $\psi$, along with their respective momenta $P_\phi$ and $P_\psi$ as canonical pairs. In terms of these variables, the constraints take the following form (setting $c=1$).
\begin{equation}
    H_T=\frac{1}{G}\int dx (N^x H_x+NH)
\end{equation}
with the constraints
\begin{align}
    H_x&=(E^x)'K_x-E^\varphi(K_\varphi)'-8\pi P_\phi\phi'-8\pi P_\psi\psi'\\
    H&=-\frac{E^\varphi}{2\sqrt{E^x}}-2\sqrt{E^x}K_\varphi K_x -\frac{K_\varphi^2 E^\varphi}{2\sqrt{E^x}}+\frac{\left((E^x)'\right)^2}{8\sqrt{E^x}E^\varphi}-\frac{2\sqrt{E^x}(E^x)'(E^\varphi)'}{2(E^\varphi)^2}+\frac{\sqrt{E^x}(E^x)''}{2E^\varphi}+\nonumber\\
    &+\frac{2\pi GP_\phi^2}{\sqrt{E^x}E^\varphi}+\frac{2\pi G\sqrt{E^x}E^x(\phi')^2}{E^\varphi}+\frac{2\pi GP_\psi^2}{\sqrt{E^x}E^\varphi}+\frac{2\pi G\sqrt{E^x}E^x(\psi')^2}{E^\varphi}.
\end{align}
Here, $x$ is a radial coordinate extended to $\mathbb{R}$ and the prime symbol denotes the derivative with respect to $x$. The algebra of constraints can be Abelianized by a change of the lapse and shift,
given by
\begin{equation}
    N^r\equiv N^x -N\frac{2\sqrt{E^x}K_\varphi}{(E^x)'},\qquad N_{\mathrm{new}}\equiv N\frac{E^\varphi}{(E^x)'}.
\end{equation}
Together with the gauge fixing condition $E^x=E^x(x)$ (independent of time), which fixes the value of $K_x$ and imposes the condition $N^x=0$, we obtain the Hamiltonian constraint
{\small{\begin{equation}\label{eq:H_Total}
    H_T=\frac{1}{G}\int dx N_{\mathrm{new}}\left[C'+
    \frac{2\pi}{\sqrt{E^x}(E^\varphi)^2}\left((E^x)'\left[((E^x)^2\left[(\phi')^2+(\psi')^2\right]+P_\phi^2+P_\psi^2\right]-8E^xE^\varphi K_\varphi\left(\phi'P_\phi+\psi'P_\psi\right)\right)\right],
\end{equation}}}
where
\begin{equation}
    C=-\sqrt{E^x}\left(1+K_\varphi^2-\frac{((E^x)')^2}{4(E^\varphi)^2}\right)+2GM.
\end{equation}
Note that the last term in $C$ encodes the ADM mass of the system and that, in vacuum, the Hamiltonian constraint implies $C'=0$. The choice of the integration constant as the mass $M$ ensures that $C=0$.

A second gauge-fixing condition, $\phi=t/L_0^2$, represents the choice of this scalar field as a clock and introduces a new length scale $L_0$, which we identify with the spatial region where the clock is synchronized. This gauge condition also fixes the momentum $P_\phi$ as well as the lapse function $N_\mathrm{new}$. The variables $E^\varphi$ and $K_\varphi$, together with the canonical pair of the remaining scalar field, constitute the (local) dynamical degrees of freedom, whose evolution is governed by the \textit{true} Hamiltonian (see details in \cite{Gambini:2023pox})
\begin{equation}
    H_\mathrm{True}=M+\int dx\frac{\sqrt{-C'(E^\varphi)^2\sqrt{E^x}-2\pi G\left((E^x)'(E^x)^2(\psi')^2-8E^xK_\varphi E^\varphi P_\psi \psi'+(E^x)'P_\psi^2\right)}}{L_0^2\sqrt{2\pi G\sqrt{(E^x)'}}}.
\end{equation}
Here, the positivity of the square root argument must be ensured by imposing suitable initial conditions on the classical solutions.

Before proceeding to quantization, we simplify the problem by retaining only the gravitational part of the true Hamiltonian---specifically, the leading-order term in the $G$-expansion of the square root. This leads to
\begin{equation}\label{eq:classical_true_hamiltonian}
    H_\mathrm{Grav}=M+\int dx\frac{\sqrt{-C'(E^\varphi)^2\sqrt{E^x}}}{L_0^2\sqrt{2\pi G\sqrt{(E^x)'}}}.
\end{equation}
Although the motivation for this work stems from resolving the quantization ambiguities usually found in gauge fixed models that were identified in a previous study of the gravitational system coupled to a clock and matter in the asymptotic approximation, we begin by analyzing these ambiguities in the vacuum case. We then prove the consistency of the quantization with the aim of extending the resolution to the coupled system throughout the entire exterior region of the black hole.

\section{Quantum theory: the Hamiltonian for gravity plus a clock}\label{sec:quantum_theory}
In this section, we present the LQG quantization of the previously described system, and study the resulting Hamiltonian. The LQG kinematical Hilbert space is given by the tensor product of radial holonomies with the Bohr compactification of the real line in the transverse direction (see details in \cite{Gambini:2022hxr}). Schematically, the spin-network states are
\[
\big|\vec{k}, \vec{\mu}\big\rangle =
\begin{array}{cccccccc}
\cdots & \overset{\mu_{j-1}}{\bullet} & - & k_j & - &
\overset{\mu_j}{\bullet} & - & k_{j+1} - \overset{\mu_{j+1}}{\bullet} \cdots
\end{array}
\]
where \(k_j \in \mathbb{Z}\) are the colors of the links and \(\mu_j \in \mathbb{R}\) are the parameters associated with the vertices. Specifically, these states are the common eigenbasis of the operators $E_j^x\equiv E_j^x(x_j)$ and $E_j^\varphi\equiv E^\varphi(x_j)$ where $x_j$ is the evaluation of $x$ at node $j$ and they are given by
\begin{eqnarray}
\hat{E}^x_{j}\left|k_1,..,k_N,\mu_0,..,\mu_{N} \right\rangle&=&k_j \ell_{Pl}^2\left|k_1,..,k_N,\mu_0,..,\mu_{N}\right\rangle\\
\hat{E}^\varphi_{j}\left|k_1,..,k_N,\mu_0,..,\mu_{N} \right\rangle&=&\mu_j \ell_{Pl}\left|k_1,..,k_N,\mu_0,..,\mu_{N} \right\rangle.\label{eq:E_phi_action}
\end{eqnarray}
To simplify the calculation, we chose the particular gauge in which $E^x(x)=x^2$ and therefore $k_j=x_j^2/\ell_{Pl}^2$. We also chose an equally spaced grid with spacing $\Delta$ (integer multiple of $\ell_{Pl}$) and a starting point $x_0$ (also an integer multiple of $\ell_{Pl}$) such that $x_j=j\Delta +x_0$. The grid goes from $x_0>R_S\equiv 2GM$ to $x_0+N\Delta$ and we identify $N\Delta=L_0$. The $\mu^0$ polymerization of the transverse direction is performed with a parameter $\rho$ independent of $x$ such that discrete translations are generated by
\begin{equation}
\exp(i\rho\hat{K}_{\varphi,j})\left|\vec{k},\mu_1,..,\mu_N \right\rangle=\left|\vec{k},\mu_0,..,\mu_j+\rho,..\mu_N\right\rangle. \end{equation}
In this kinematical space, we introduce the Hamiltonian operator corresponding to $H_\mathrm{Grav}$ in the next subsection.
The previous discussion omitted the sector of the Hilbert space that accounts for global degrees of freedom, that is, the ADM mass, which results from imposing asymptotic conditions. For simplicity, we assume that the system is an eigenstate of mass with eigenvalue $M$ and omit the notation related to that sector of the state space.

\subsection{The Hamiltonian}
The Hamiltonian in (\ref{eq:classical_true_hamiltonian}) needs to be modified before quantization to ensure its self-adjointness. The extension
\begin{equation}\label{eq:Hamiltonian_grav}
H=M+\int dx\frac{\left| E^\varphi (x)\right|\sqrt{\left|C'(x)\right|\sqrt{\left|E^x(x)\right|}}}{L_0^2\sqrt{2\pi G (E^x(x))'}}
\end{equation}
suffices and reduces to $H_\mathrm{Grav}$ in the classical solutions (such that $C'\leq 0$).
After symmetrization (also required for selfadjointness), on the kinematical space described before, we get the Hamiltonian
\begin{equation}\label{eq:Hamiltonian}
    \hat{H}=M+\sum_{j=0}^{N-1}\frac{F\Delta}{2}\left[|E_j^\varphi|\sqrt{\frac{|\hat{C}_{j+1}-\hat{C}_j|}{\Delta}}+\sqrt{\frac{|\hat{C}_{j+1}-\hat{C}_j|}{\Delta}}|E_j^\varphi|\right],
\end{equation}
where $F=(\sqrt{4\pi}\ell_{Pl}L_0^2)^{-1}$ and $C'\to\frac{C_{j+1}-C_j}{\Delta}$. The constraint function $C$ has been replaced by
\begin{equation}
\hat{C}_j=-\sqrt{\hat{E}^x_{j}}\left[1+\frac{\sin\left(\rho\hat{K}_{\varphi,j}\right)^2}{\rho^2}-\frac{\left(\hat{E}^x_{j+1}-\hat{E}^x_{j}\right)^2}{4\Delta^2\left(E_j^\varphi\right)^2}\right]+R_S
\end{equation}
Thiemann's trick is now used to extend the operator $\left(E_j^\varphi\right)^{-2}$ to $\mu=0$. This is done using the definition:
\begin{equation}
\left(E_j^\varphi\right)^{-2}\left|\vec{k},\vec{\mu}\right\rangle=\left(\frac{3}{4\rho}\right)^6\frac{1}{\ell_{Pl}^2}\left(\left|\mu_j+\rho\right|^{2/3}-\left|\mu_j-\rho\right|^{2/3}\right)^6\left|\vec{k},\vec{\mu}\right\rangle
\end{equation}
and $1/(\mu_j)^2$ is recovered for a sufficiently small polymerization parameter $\rho$. As can be seen in (\ref{eq:Hamiltonian}), studying $\hat{H}$ requires knowledge of the spectrum of $\hat{C}_j$. This will be the subject of the next section.
Note that in previous studies \cite{Gambini:2023pox,Gambini:2024suj,Gambini:2023wnc}, similar expressions for $\hat{C}_j$ include a deficit angle that does not appear here because we neglect the contribution of matter fields other than the clock.

\subsection{Finite Difference equation for the Hamiltonian constraint}\label{sec:Finnite_Diff}

As mentioned previously, our goal was to study the spectrum of the Hamiltonian $\hat{H}$. In particular, we aim to show that the solutions of the pure-gravity quantum Hamiltonian constraint are recovered at the limit where the clock mass becomes negligible. As can be seen from (\ref{eq:Hamiltonian_grav}), for clocks of negligible energy one has $C'=0$ and $P_\phi=0$, which implies $H_\mathrm{Grav}=M$. Therefore, we must verify that the gauge-fixed quantum Hamiltonian has eigenvalues sufficiently close to $M$ to recover the solutions of the pure-gravity Hamiltonian constraint.
As a preliminary step towards analyzing the Hamiltonian spectrum, we considered the following eigenvalue equation:
\begin{equation}
    \hat{C}_j\left|\phi_j\right\rangle=l_j\left|\phi_j\right\rangle.
\end{equation}
On the $\left|\vec{k},\vec{\mu}\right\rangle$ basis, the $\hat{C}_j$ operator acts trivially on all components except $\mu_j$, and the equation can be written as
\begin{equation}
-\phi_j(\mu+2\rho)-\phi_j(\mu-2\rho)+2\phi_j(\mu)+V_j(\mu)\phi_j(\mu)=\lambda_{j}\phi_j(\mu)\label{eq_discreta}
\end{equation}
where 
\begin{eqnarray}
\phi_j(\mu)&\equiv&\left\langle\vec{k},\mu_1,..\mu_{j-1},\mu,\mu_{j+1},..\mu_N\right|\phi_j\rangle,\\
V_j(\mu)&\equiv&-\rho^2\left(\frac{2x_j+\Delta}{\ell_{Pl}}\right)^2\left(\frac{3}{4\rho}\right)^6\left(\left|\mu+\rho\right|^{2/3}-\left|\mu-\rho\right|^{2/3}\right)^6\\
\lambda_j&=&\lambda(l_j,x_j)\equiv 4\rho^2\left[\frac{R_S-l_j}{|x_j|}-1\right]\label{eq:lambda_def}.
\end{eqnarray}
For convenience, we introduce the function $\lambda$, which maps the eigenvalues $l_j$
 of $\hat{C}_j$ at $x_j$ to those appearing in (\ref{eq_discreta}). By setting $\mu=2\rho n +\epsilon$, with $0<\epsilon< \rho$ identifying the superselection sector on which the theory is defined, equation (\ref{eq_discreta}) reduces to a discrete version of the time-independent Schrödinger equation
\begin{equation}\label{eq:diff_finita}-\Delta^2\Phi_{\lambda_j}(n-1)+U_j(n)\Phi_{\lambda_j}(n)=\lambda_j\Phi_{\lambda_j},
\end{equation}
where $\Delta$ is the forward difference operator $\left[\Delta\phi(n)=\phi(n+1)-\phi(n)\right]$, $\Phi_{\lambda_j}(n)\equiv\phi_j(2\rho n +\epsilon)$ is a redefinition of the eigenfunction and
\begin{equation}\label{potential}
U_j(n)\equiv V_j(2\rho n +\epsilon)=-\left(\frac{2x_j+\Delta}{\ell_{Pl}}\right)^2\left(\frac{3}{4}\right)^6\left(\left|2n+1+\frac{\epsilon}{\rho}\right|^{2/3}-\left|2n-1+\frac{\epsilon}{\rho}\right|^{2/3}\right)^6.
\end{equation}
In what follows, we refer $U_j$ as \textit{the potential} in analogy with the time-independent Schrödinger problem.
\subsection{Spectral analysis}\label{sec:spectral_analysis}
The spectrum of the \textit{Schrödinger operator} $-\Delta^2+U(n)$ for $n\in\ell^2(\mathbb{Z})$ has been studied in the literature \cite{damanik_teschl_2007,Damanik2004HalflineSO,Bach2018,Kim2009} for a variety of potentials $U$. In cases where $U(n)\to0$ when $n\to\pm\infty$, it features a continuous or \textit{essential} spectrum with eigenvalues in the $\left[0,4\right]$ interval and there is room for a discrete spectrum of normalizable eigenfunctions depending on the potential. If it is bounded (as in our case, due to Thiemann's trick), the complete spectrum is also bounded. For strictly negative potential \textit{wells}, 
like the one analyzed here, the discrete spectrum is found below the bottom of the essential spectrum (i.e., with eigenvalues smaller than 0), and its nature is determined by the $n\to\pm\infty$ asymptotics of the potential. For example, for wells with $U\sim n^{-2+\delta}$, the discrete spectrum is finite for $\delta<0$ and infinite for $\delta>0$. The asymptotics of interest correspond to the critical value $\delta =0$ which is a more delicate case, but some general properties are known. In particular, if we write (\ref{potential}) in the form
\begin{equation}
U_j(n)=-\frac{\alpha_j^2}{n^2}+W(n)
\end{equation}
where $\alpha_j=\frac{|2x_j+\Delta|}{2\ell_{Pl}}$, $W(n)=O(n^{-4})$ and take into account that the (\textit{supercritical}) condition $\alpha_j>1/2$ is satisfied, the eigenvalue problem (\ref{eq:diff_finita}) falls under the hypothesis of Theorem 1 in \cite{damanik_teschl_2007}. There, the following is proven:

\noindent\textit{The discrete spectrum of (\ref{eq:diff_finita}) has an accumulation point at $\lambda_j=0$ and the number $N(E)$ of eigenvalues below $E<0$ has the property\footnote{Notice that (\ref{eq:lim_spectrum}) is twice the result of \cite{damanik_teschl_2007}. The reason being that they consider the eigenvalue problem with vanishing Dirichlet conditions for $n\in\mathbb{Z}^+$ and we consider $n\in\mathbb{Z}$. For the particular case $\epsilon=0$ our solutions split into even and odd families and their problem corresponds to the odd solutions. For a general case (with small $\epsilon$) the conclusion is the same by continuity with respect to the parameter.}
    \begin{equation}\label{eq:lim_spectrum}
        \lim_{E\to 0^-}\frac{N(E)}{-\ln(-E)}=\frac{1}{\pi}\sqrt{\alpha_j^2-\frac{1}{4}}.
    \end{equation}}

In summary, for $\lambda_j<0$ we have a bounded discrete spectrum of normalizable eigenfunctions with an accumulation point at $\lambda_j=0$, followed by a continuous spectrum for $\lambda_j\in(0,4)$. 

\subsubsection{Uniform WKB approximation}\label{sec:WKB}
We do not have access to the exact solutions; however, we can obtain an approximation for the spectrum of (\ref{eq:diff_finita}). We are primarily interested in the discrete spectrum ($\lambda_j<0$), where the solutions correspond to the bound states in a potential well with two \textit{classical} turning points ($n_\pm$) given by the condition $U_j(n_\pm)=\lambda_j$. Given that the potential has an overall factor of $\alpha_j^2\sim |x_j|^2/\ell_{Pl}^2$, -a huge number of macroscopic black holes- and that $\lambda_j\propto-4\rho^2$ according to (\ref{eq:lambda_def}), which is a scale of the system not fixed a priori, it is possible to assume $|n_\pm|>>1$. This allowed us to implement several approximations. In particular, the potential can be substituted for its asymptotic form
\begin{eqnarray}
    U_j(n)&\sim&-\alpha_j^2/n^2,\qquad n\neq 0\\
    U_j(0)&=&0\nonumber,
\end{eqnarray}
which is a symmetric function with turning points $n_\pm=\pm \alpha_j/\sqrt{-\lambda_j}$. In addition, because the main contribution to the eigenfunctions comes from the vicinity of the turning points, we can approximate the finite-difference equation using a differential equation (see Appendix \ref{sec:WKB_appendix}). This simplifies the analysis and allows us to implement a uniform WKB approximation (see \cite{Ghatak2011}) that describes the eigenfunctions in the entire range of $n$. The errors coming from the continuous approximation and the small $n$ region are sub-leading and can be absorbed in a parameter, as discussed below. In Appendix \ref{sec:WKB_appendix} we find the eigenstates under a uniform WKB ansatz and obtain
\begin{equation}\label{eq:eigenfunctions}
    \Phi_{\lambda_j}(n)=C_j\frac{\mathrm{Ai}[\zeta(y)]}{\sqrt{\zeta'(y)}}
\end{equation}
where $y=n/n_+$, $\mathrm{Ai}$ is the normalizable Airy function and
\begin{equation}
\zeta(y)\equiv\left[\frac{3\alpha_j}{2}\int_1^y d\bar{y}\sqrt{1/\bar{y}^2-1}\right]^{2/3},\qquad |y|>1/n_+.
\end{equation}
To obtain well-defined solutions across the entire $n$ range, there is a consistency condition (Bohr-Sommerfield rule) given by
\begin{equation}
 2\alpha_j\int_{1/n_+}^{1}d\bar{y}\sqrt{1/\bar{y}^2-1}=\pi(\kappa+1/2).    
\end{equation}
The natural number $\kappa$ defines the spectrum levels, and the associated eigenfunctions share parity due to the term $\pi\kappa$ in the previous equation. 
Solving for $\lambda_j$ we obtain (see Appendix \ref{sec:WKB_appendix})
\begin{equation}\label{eq:lambda_spectrum}
\lambda_{j,\kappa}\equiv\lambda_{j,0}\exp\left(-\frac{\kappa\pi}{\alpha_j}\right),\quad \kappa=0,1,...
\end{equation}
where $\lambda_{j,0}$ is the ground level, and it encodes all dependences on the discreteness and shape of the potential due to Thiemman's trick. Note that, for a large $\alpha_j$, the separation between the eigenvalues is
\begin{equation}\label{eq:eigenvalue_separation}
\delta\lambda_{j,\kappa}\equiv\lambda_{j,\kappa}-\lambda_{j,\kappa}\sim-\lambda_{j,\kappa}\frac{\pi}{\alpha_j}
\end{equation}
which is an extremely small spacing. Moreover, the spectrum exhibits the expected accumulation point at $\lambda_{j,\infty}=0$. Indeed, one can readily verify that it obeys equation (\ref{eq:lim_spectrum}) for $\alpha_j>>1/4$. 
From now on, lets index the eigenfunctions and eigenvalues using an integer $\kappa$. Equation (\ref{eq:eigenfunctions}) serves as an approximation for the spectrum; however, such indexing is always possible because the discrete spectrum is bounded below and has only one accumulation point at $\lambda_j=0$. Because there are $N$ different eigenvalue problems (one for each $x_j$), we have a vector $\vec{\kappa}=(\kappa_1,..,\kappa_N)$ of indices, living in the positive cubic lattice $\mathbb{N}^{N}$. Then we define the basis $\left|\vec{\kappa}\right\rangle$ of
simultaneous eigenstates of all the $\hat{C}_j$'s given by
\begin{equation}\label{eq:kappa_basis}
    \left.\left\langle\vec{k},\vec{\mu}\right|\vec{\kappa}\right\rangle=\prod_{j=1}^N\Phi_{\lambda_{j,{\kappa_j}}},
\end{equation}
where $\Phi_{\lambda_{j,{\kappa_j}}}$ are normalized solutions to (\ref{eq:diff_finita}) with eigenvalues ${\lambda_{j,{\kappa_j}}}$. This is a key component in the next section.
\subsection{Hamiltonian}\label{sec:Hamiltonian}
Let us apply knowledge about the eigenstates of $\hat{C}_j$ to study the gravitational contribution to the true Hamiltonian given by the operator (\ref{eq:Hamiltonian}). To find its ground level, consider the basis of states $\left|\vec{\kappa}\right\rangle$ introduced in (\ref{eq:kappa_basis}). A generic normalizable state on this basis is
\begin{equation}
    \left|\Psi\right\rangle=\sum_{\vec{\kappa}}a(\vec{\kappa})\left|\vec{\kappa}\right\rangle
\end{equation}
where the sum runs through the discrete spectrum in equation (\ref{eq:diff_finita}). The continuous part of the spectrum, above $\lambda_j=0$, is ignored because it does not lead to a well-defined Hamiltonian density. The expectation value of $\hat{H}$ for such states is
{\small{\begin{equation}
    \left\langle\Psi\right|\hat{H}\left|\Psi\right\rangle=M+\sum_{j=1}^{N-1} \frac{F\Delta}{2}\sum_{\vec{\kappa},\kappa'_j}a(\vec{\kappa})a^*(\kappa_1,..,\kappa'_j,..,\kappa_N)\left\langle\kappa_j\left||E_j^\varphi|\right|\kappa_j'\right\rangle\left[\sqrt{\frac{|l_{j+1,{\kappa_{j+1}}}-l_{j.{\kappa_j}}|}{\Delta}}+\sqrt{\frac{|l_{j+1,{\kappa_{j+1}}}-l_{j,{\kappa_j'}}|}{\Delta}}\right].
\end{equation}}}%
where the bracket on the right side comes from writing $\left|\vec{\kappa}\right\rangle=\prod_j\left|\kappa_j\right\rangle$ and $l_{j,{\kappa}}$ represents the $\kappa$'th eigenvalue of $\hat{C_j}$. From inspection, we see that the absolute minimum of the sum is achieved by conditions $l_{j+1,{\kappa_{j+1}}}=l_{j,{\kappa_j}}=l_{j,{\kappa'_j}}$ for all $j$,  canceling both square roots. The second condition is met by restricting the summation to elements of the basis $\left|\vec{\kappa}\right\rangle$, states for which $\kappa_j=\kappa_j'$ for all $j$. This reduces the summation to 
\begin{equation} \left\langle\vec{\kappa}\right|\hat{H}\left|\vec{\kappa}\right\rangle=M+\sum_{j=1}^{N-1}F\Delta\left\langle\kappa_j\left||E_j^\varphi|\right|\kappa_j\right\rangle\sqrt{\frac{|l_{j+1,{\kappa_{j+1}}}-l_{j,{\kappa_j}}|}{\Delta}}.
\end{equation}
However, a sequence that fulfills $l_{j,{k_j}}=l$ for all $j$, canceling the remaining square root, is not expected from a discrete spectrum. Within the $\left|\vec{\kappa}\right\rangle$ basis we have found states that minimize the square roots while keeping the prefactor (i.e. the expectation value of $\left|E^\varphi_j\right|$) relatively small as we will shortly see.

In Appendix \ref{sec:Ephi_appendix} we compute the matrix element of $|E_j^\varphi|$ for the $\left|\vec{\kappa}\right\rangle$ states. In particular, assuming $|x_j|/\ell_{Pl}\sim\alpha_j>>1$ we obtain the following
\begin{equation}\label{eq:E_expectation_value}
\left\langle\kappa\left||E_j^\varphi|\right|\kappa\right\rangle\sim\frac{|x_j|}{\sqrt{1-\frac{R_S-l_{j,{\kappa}}}{|x_j|}}}\frac{\pi}{4},
\end{equation}
which provides a finite result for states away from\footnote{This condition identifies highly exited clock states, as we will discuss later.} $R_S-l_{j,{\kappa}}=|x_j|$. This simplifies the expectation value of $\hat{H}$ to
\begin{equation} \label{eq:energy}\left\langle\vec{\kappa}\right|\hat{H}\left|\vec{\kappa}\right\rangle=M+\sum_{j=1}^{N-1}F\Delta\frac{\pi}{4}\frac{|x_j|}{\sqrt{1-\frac{R_S-l_{j,{\kappa_j}}}{|x_j|}}}\sqrt{\frac{|l_{j+1,{\kappa_{j+1}}}-l_{j,{\kappa_j}}|}{\Delta}}.
\end{equation}

\begin{figure}
    \centering
\includegraphics[width=0.5\linewidth]{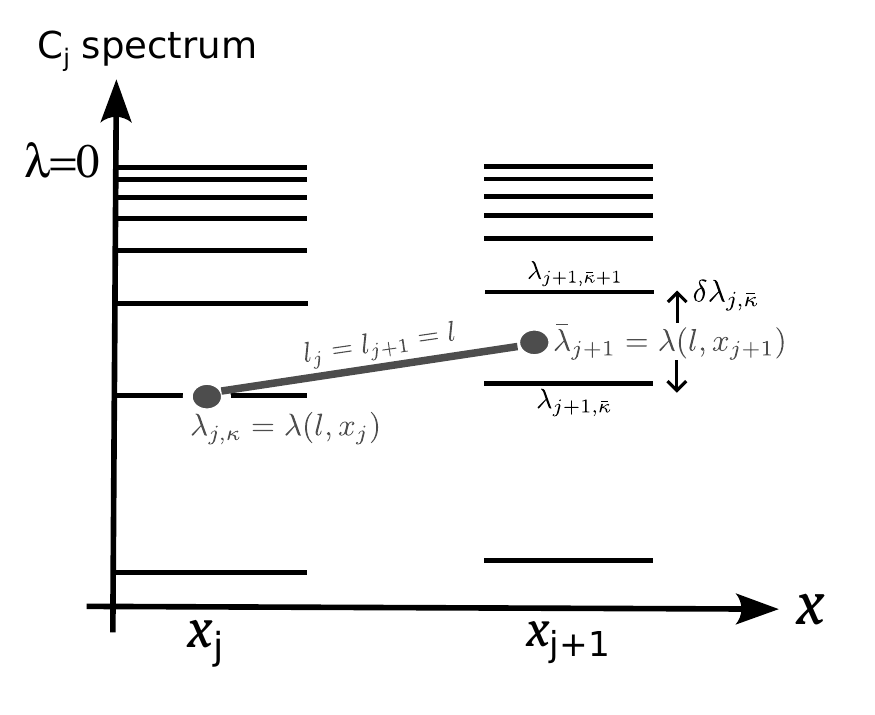}
    \caption{Scheme on how to approximate $\delta l_j=|l_{j,\kappa_j}-l_{j+1,\kappa_{j+1}}|$ in equation (\ref{eq:energy}). The best approximation to $l_{j+1}=l_j=l$ is provided by the closest eigenvalues to $\bar{\lambda}_{j+1}=\lambda(l,x_{j+1})$, that is, $\lambda_{j+1,\bar{\kappa}}$ and $\lambda_{j+1,\bar{\kappa}+1}$. In particular, $|l_{j+1,\bar{\kappa}+1}-l_{j+1,\bar{\kappa}}|$ serves as an upper bound for $\delta l_j$ and is proportional to $\delta\lambda_{j,\bar{\kappa}}$, for which we have the explicit expression in (\ref{eq:eigenvalue_separation}).}
    \label{fig:approx_spectrum}
\end{figure}

To find the minimum of the square root, we can do the following estimation. Suppose $l_{j,{\kappa}}=l$. Then, according to (\ref{eq:lambda_def}),
\begin{equation}
    \lambda_{j,\kappa}\equiv\lambda(l,x_j)=4\rho^2\left(\frac{R_S-l}{|x_j|}-1\right).
\end{equation}
The corresponding $\lambda_{j+1}$ to cancel the square root would be
\begin{equation}
    \bar{\lambda}_{j+1}\equiv\lambda(l,x_{j+1})=4\rho^2\left(\frac{R_S-l}{|x_{j+1}|}-1\right).
\end{equation}
However, this is likely not part of the spectrum of $\hat{C}_{j+1}$. Let us name the eigenvalues closest to $\bar{\lambda}_{j+1}$ as $\lambda_{{j+1},\bar{\kappa}}$ and $\lambda_{{j+1},{\bar{\kappa}+1}}$, then a good approximation for $\delta l_j\equiv l_{j+1,{\kappa_{j+1}}}-l_{j,{\kappa_j}}$ is (see figure \ref{fig:approx_spectrum})
\begin{equation}\label{eq:delta_l}\delta l_j\sim\left|l_{j+1,{\bar{\kappa}+1}}-l_{j+1,{\bar{\kappa}}}\right|=\frac{|x_{j+1}|}{4\rho^2}\left|\lambda_{j+1,{\bar{\kappa}+1}}-\lambda_{j+1,\bar{\kappa}}\right|\sim-\frac{\bar{\lambda}_{j+1}}{4\rho^2\pi}\ell_{Pl}=\left(1-\frac{R_s-l}{|x_{j+1}|}\right)\frac{\ell_{Pl}}{\pi},
\end{equation}
where approximation (\ref{eq:eigenvalue_separation}) and equation (\ref{eq:lambda_def}) are used in the middle step. By substituting into (\ref{eq:energy}), we obtain
\begin{equation}\label{eq:E_discrete}
\left\langle\vec{\kappa}\right|\hat{H}\left|\vec{\kappa}\right\rangle=M+\sum_{j=1}^{N-1}F\Delta\frac{\sqrt{\pi}}{4}|x_j|\sqrt{\left|\frac{|x_{j+1}|-R_S+l}{|x_j|-R_S+l}\right|\frac{|x_j|}{|x_{j+1}|}}\sqrt{\frac{\ell_{Pl}}{\Delta}}\sim M+\sum_{j=1}^{N-1}F\Delta\frac{\sqrt{\pi}}{4}|x_j|\sqrt{\frac{\ell_{Pl}}{\Delta}}.
\end{equation}
Here, we assume that $x_j$ is sufficiently large to approximate the square roots in the middle steps to one. Finally, assuming $\ell_{Pl}=\Delta$ and returning the expression for $F$ we find that
\begin{equation}
\left\langle\vec{\kappa}\right|\hat{H}\left|\vec{\kappa}\right\rangle\sim M+M_{Pl}\frac{\sum_{j=1}^{N-1}|x_j|\Delta}{8L_0^2}=M+\mathcal{O}(M_{Pl}).
\end{equation}
Note that we obtained an upper bound on the ground state energy, but this result also shows that the particular state used in the computation provides a good approximation to the ground state.

Having described the ground state, we may ask whether the same basis of eigestates of $\hat{C}_j$ is useful for describing a highly excited clock field. To address this, we return to equation (\ref{eq:energy}) and consider the limit in which all eigenvalues $ \lambda_{j,\kappa_j}$ coincide and approach the accumulation point $\lambda_{j,\infty}=0$, instead of minimizing $|l_{j+1,\kappa_{j+1}}-l_{j,{\kappa_j}}|$. These states correspond to $l_{j,\kappa_j}\to l_{j,\infty}= R_S-|x_j|$ which implies that $|l_{j+1,\kappa_{j+1}}-l_{j,{\kappa_j}}|\to|l_{j+1,\infty}-l_{j,\infty}|=\Delta$. Then, in that limit
\begin{equation}
\left\langle\hat{H}\right\rangle\sim M+\sum_{j=1}^{N-1}\frac{F\Delta|x_j|^{3/2}}{\sqrt{l_{j,\kappa_j}-R_S+|x_j|}}.
\end{equation}
The energy density becomes arbitrarily large as $l_{j,{\kappa_j}}\to R_S-|x_j|$ ($\kappa_j\to+\infty$). Therefore, the discrete spectrum of $\hat{C}_j$ is sufficient to characterize the entire spectrum of $\hat{H}$.\\
In summary, we have found that the ground state can be found within the discrete basis $\left|\vec{\kappa}\right\rangle$ of simultaneous eigenstates of all $\hat{C}_j$ operators, with eigenvalue of order $M+M_{Pl}$. We also find that this basis proves the entire spectrum of the Hamiltonian.

\section{Analytical approximation to the spectrum}\label{sec:Analytic_approx}
Consider expression (\ref{eq:lambda_spectrum}) valid for the entire discrete spectrum. Subsequently, if we substitute it directly into (\ref{eq:energy}), we obtain the following expression for the expectation value of $\hat{H}$
\begin{equation}
\begin{aligned}
\langle \vec{\kappa} | \hat{H} | \vec{\kappa} \rangle
= M
+ M_{Pl}
\sum_{j=1}^{N-1}
\frac{\,\Delta |x_j|}{\sqrt{4\pi}L_0^2}
\exp\!\left(\frac{\pi\kappa_j\ell_{Pl}}{2|x_j|}\right)
\sqrt{\Xi_j} .
\end{aligned}
\end{equation}
where
\begin{equation}
\Xi_j \equiv
\left|
\frac{|x_j|}{\Delta}
\exp\!\left(-\frac{\pi\kappa_j\ell_{Pl}}{|x_j|}\right)
-
\frac{|x_{j+1}|}{\Delta}
\frac{\lambda_{j+1,0}}{\lambda_{j,0}}
\exp\!\left(-\frac{\pi\kappa_{j+1}\ell_{Pl}}{|x_{j+1}|}\right)
+1\right|.
\end{equation}
Notice that the Hamiltonian density exhibits exponential growth for $\kappa_j>>
|x_j|/\ell_{Pl}$ because in this regime $\Xi_j\to1$. In contrast, for fixed $|\vec{\kappa}|<<\frac{|x_j|}{\ell_{Pl}}$ the expression reduces to (\ref{eq:Hamiltonian_small_kappa}), yielding Planck-scale energy spacings between neighboring eigenvalues.
\begin{equation}\label{eq:Hamiltonian_small_kappa}
\left\langle\vec{\kappa}\right|\hat{H}\left|\vec{\kappa}\right\rangle\sim M+M_{Pl}\sum_{j=1}^{N-1}\frac{\Delta|x_j|}{\sqrt{4\pi}L_0^2}\sqrt{\left|1+\left(\kappa_{j+1}\frac{\lambda_{j+1,0}}{\lambda_{j,0}}-\kappa_j\right)\frac{\pi\ell_{Pl}}{\Delta}\right|}.  
 \end{equation}
Because this spectrum originates from a WKB analysis, corrections are expected for lower $\kappa$'s. However, given that $\lambda_{j,\infty}=0$ is the only accumulation point and corresponds to the higher $\kappa$'s, this analysis could be complemented with a numerical treatment for the (finite) lower part of the spectrum of $\hat{C}_j$ to improve the approximations considered in this and previous sections. 

\section{Comparison with the \texorpdfstring{$\rho\to0$}{rho to 0} limit.}\label{sec:comparison}
In a previous study \cite{Gambini:2023wnc}, this system was analyzed in the continuous limit $\rho\to0$. In this regime, equation (\ref{eq_discreta}) becomes a differential equation:
\begin{equation}\label{eq:cont_eigenvalue_problem}
    -\frac{d^2}{d\mu^2}\phi_j(\mu)+V_j(\mu)\phi_j(\mu)=-l_x\phi_j(\mu)
\end{equation}
with potential $V_j(\mu)=-\alpha^2_j/\mu^2$ and $l_x=\frac{l_j-R_S}{|x_j|}+1$. It is well known (see Section 35 of \cite{LandauLifshitzQM}) that this potential exhibits a \textit{fall to the center} phenomenon at $\mu=0$.  Consequently, the discrete spectrum is unbounded from below, requiring a renormalization criterion to properly recover the eigenvalues and eigenfunctions. In the absence of such a criterion, only the asymptotic region ($\mu>>1$) can be accessed, where the equation admits exact solutions but lacks a discretization condition for the spectrum. Consequently, the spectrum was treated as continuous in the cited study. Although the discrete spectrum was treated as a continuum, the condition $\lambda_j<0$ remained, because it is required for normalizability, and it reads
\begin{equation}
    l_j> R_S-|x_j|.
\end{equation}
Under that assumption, the eigenfunctions take the asymptotic form
\begin{equation}
    \phi_j(\mu)\propto \exp\left[\pm \mu\sqrt{\frac{l_j+|x_j|-R_S}{|x_j|}}\right],
\end{equation}
where the normalizable ones correspond to the decaying exponentials in the positive and negative regions of the variable. This is precisely the limit of the eigenfunctions (\ref{eq:eigenfunctions}) as $\rho\to0$ and $|n|>>n_+$ with $\mu=2\rho n+\epsilon$ held constant. A key point of the cited work is that, within a continuous spectrum, the ground state can be represented by a sequence of solutions to (\ref{eq:cont_eigenvalue_problem}) with constant eigenvalues $l=l_1=l_2=...$, but the resulting state becomes non-normalizable. Then, a superposition of eigenstates is required to represent the physical states and ansatz
\begin{equation}
    \Delta l_j=\frac{x_j^2}{\Delta}\left(\frac{\Delta}{L_0}\right)^{2a}
\end{equation}
is used in \cite{Gambini:2023wnc} for the standard deviation of the wave-packets around an arbitrary eigenvalue $l$ of $\hat{C}_j$. In analogy to equation (\ref{eq:E_expectation_value}), the expectation value of $\left|E^\varphi_j\right|$ is calculated with the result
\begin{equation}
\left\langle|E_j^\varphi|\right\rangle=\left(1-\frac{R_S-l}{|x_j|}\right)^{-1/2}|x_j|.
\end{equation}
This is important because, in the classical vacuum solution, one has
\begin{equation}
    |E^\varphi|=\sqrt{g_{xx}|E^x|}= |x|/\sqrt{1-R_S/|x|},
\end{equation}
implying $l=0$ to recover the correct asymptotics for this expectation value. In our calculation there is a pre-factor $\pi/4$. However, using the result (\ref{eq:E_matrix_element}) from Appendix \ref{sec:WKB_appendix}, we have that
\begin{equation}
\left\langle\kappa\left||E_j^\varphi|\right|\kappa+2\right\rangle\sim\frac{|x_j|}{\left(1-\frac{R_S-l_{j,\kappa}}{|x_j|}\right)^{1/4}\left(1-\frac{R_S-l_{j,\kappa+2}}{|x_j|}\right)^{1/4}}(-\frac{\pi}{4}).
\end{equation}
Given that $l_{j,\kappa}$ and $l_{j,\kappa+2}$ are extremely close (separated by a distance of order Planck), we can chose the closest eigenvalue to $l_{j}=0$ (let say $l_{j,\kappa_j}\sim0$) for each $j$ and do the combination $\left|\Psi\right\rangle=\frac{1}{\sqrt{2}}\left[\left|\vec{\kappa}\right\rangle-\left|\vec{\kappa}+\vec{2}\right\rangle\right]$, where all $\kappa_j$ are shifted by two. That way we obtain
\begin{equation}
\left\langle\Psi\left||E_j^\varphi|\right|\Psi\right\rangle\sim\frac{|x_j|}{\sqrt{1-\frac{R_S}{|x_j|}}},
\end{equation}
with corrections of order $\ell_{Pl}/|x_j|$. Since the eigenvalues in the combination are so close to each other, none of the qualitative features of the states $\left|\vec{\kappa}\right\rangle$ is lost if substituted by $\left|\Psi\right\rangle$. In particular, all conclusions from the previous section apply if we switch to these states with the correct asymptotics. It is worth mentioning that such $\left|\Psi\right\rangle$ states have large uncertainty. This indicates that the ground state of $\hat{H}$ is far from semi-classical.

An other aspect of our calculation, absent in the continuum limit is that, according to (\ref{eq:delta_l}), we have $l_j$ jumping to
\begin{equation}\label{eq:random_jump}
    l_{j+1}=l_j\pm\left(1-\frac{R_S-l_j}{|x_j|}\right)\pi\ell_{Pl}
\end{equation}
with the sign changing almost randomly. Therefore, the expectation value of this distribution should be $<l_j>=0$ to recover the asymptotic form of the metric. The specifics of the standard deviation depend on the details of this randomness but is safe to assume from equation (\ref{eq:random_jump}) that it is of order $\ell_{Pl}$. Taking this as uncertenty in $l_j$ and comparing with the referred ansatz we see that, in the asymptotic limit ($x_j\sim L_0$) the ansatz reduces to $\Delta^{2a-1}/L_0^{2a-2}$, which is of order Planck if $a=1$ and $\Delta\sim\ell_{Pl}$. The treatment of the matter Hamiltonian has not been the subject of the present work, but  this value of parameter $a$, as shown in \cite{Gambini:2023wnc}, fixes ambiguities in the asymptotic quantization of the matter field coupled to gravity and allows the recovery of the expected form of the matter Hamiltonian.

Finally, constraint $C'=0$ is achieved as the expectation value of $<\frac{l_{j+1}-l_j}{\Delta}>=0$ for this random distribution.
\section{Conclusions}
In this study, we established rigorous foundations for the quantization of gravity coupled to a scalar material field outside the event horizon of a black hole. This is a problem that has remained unsolved, until now, because of the impossibility of finding quantum constraints in LQG that reproduce the classical algebra. In a previous study, we considered the gauge-fixed quantization of the system using a second scalar field as a clock, and restricted the analysis to the asymptotic region. This forced us to make several ad hoc approximations of the theory, preventing a rigorous analysis of the consistency and uniqueness of the quantization. First, we assumed an approximation of the true Hamiltonian of the gauge-fixed quantum theory with a vanishing polimerization parameter. This hypothesis simplifies the problem of determining the spectrum of the Hamiltonian by allowing the use of differential equations instead of finite difference equations. However, it eliminates the discrete normalizable sector of solutions, forcing the introduction of wave packets whose properties must be determined guided by considerations of consistency with the usual field theory of the matter field in the asymptotic limit. Here, we study the gauge-fixed $\mu_0$ loop quantization of the system describing the entire exterior region of a black hole, rather than restricting the analysis to the asymptotic regime. This requires a treatment where the asymptotic approximations considered in previous papers do not hold, and careful control of both the consistency of the gauge-fixed formulation and the identification of the correct factor ordering. In this work, we examine these issues by determining whether the gauge-fixed quantization reproduces the established results obtained from the Dirac quantization of a vacuum black hole. We first concentrate on the purely gravitational sector, assuming no matter fields other than the clock. Within this framework, we resolve the ambiguities associated with defining the spectrum of the true Hamiltonian, check its consistency with the Hamiltonian constraint, and establish the basis for a future numerical investigation of the exact spectrum and quantum effects arising when matter is coupled to gravity. Finally, we analyzed the validity of some of the assumptions made in \cite{Gambini:2023wnc}, where the total interacting system was studied under more restrictive approximations. The groundwork is thus laid for a rigorous and fully quantum study of matter coupled with gravity in spherically symmetric LQG.
\section*{Acknowledgment}
This work was supported in part by PEDECIBA and ANII (Fondo Clemente Estable FCE-1-2023-1-175902).
\appendix
\section{Uniform WKB approximation}\label{sec:WKB_appendix}
In this section, we compute the eigenfunctions and eigenvalues of the discrete spectrum in equation (\ref{eq:diff_finita}). We are not aware of an exact solution to this spectrum in terms of known special functions and, in fact, one does not expect to have them for any potential other than very specific examples. However, WKB-like methods provide a good approximation of the spectrum and eigenfunctions. The situation is analogous to the continuous case, where the Schrödinger equation is known to be well approximated by oscillatory functions in classically available regions and by real exponentials in classically forbidden regions. The matching conditions at the interfaces can be presented as Bohr-Sommerfeld rules for the bound states, and lead to approximated expressions for the discrete spectrum. In our case, the bound states correspond to $\lambda_j<0$ and the classically allowed regions are bounded by two turning points ($n_\pm$) given by $U_j(n_\pm)=\lambda_j$. As is, the implementation of a WKB ansatz is possible, but cumbersome. However, we can take advantage of two facts with regard to our problem. The potential is proportional to $\alpha_j^2\sim|x_j|^2/\ell^2_{Pl}$ according to (\ref{potential}), which is a huge number for a macroscopic black hole. In addition, $\lambda_j\propto-\rho^2$ with $\rho$ the polymerization parameter, whose scale is not fixed a priori. Taking $\rho<<|x_j|/\ell_{Pl}$, which is not particularly restrictive, we ensure that the turning points occur at large $n$. This allows for the implementation of two main approximations. We use the asymptotic form of the potential
\begin{eqnarray}
    U_j(n)&\sim&-\alpha_j^2/n^2,\qquad n\neq 0\\
    U_j(0)&=&0\nonumber,
\end{eqnarray}
which is a symmetric function with turning points $n_\pm=\pm \alpha_j/\sqrt{-\lambda_j}$, and we approximate the discrete variable $n$ by a continuous variable $y=n/n_+$. This is possible because the eigenfunctions have their main contributions from the vicinity of the turning points, where the increment in $n$ is much smaller than the value of $n$ itself. This allows us to treat equation (\ref{eq:diff_finita}) as a differential equation
\begin{align}\label{eq:dif_equation}
    &\frac{d^2\Phi_j(y)}{dy^2}+\alpha_j^2[F(y)-1]\Phi_j(y)=0,\\
    &F(y)=1/y^2,\quad |y|\geq1/n_+\\
    &F(y)=0,\qquad |y|<1/n_+
\end{align}
and use a uniform version of the WKB method introduced by Langer \cite{Langer1937} and pedagogically explained in \cite{Ghatak2011}. First, we introduce the function
\begin{align}
\Gamma_j^2(y)&=\alpha_j^2[F(y)-1]
\end{align}
and the change of variable
\begin{equation}\label{eq:def_zeta}
    \zeta_j(y)=\left(\int_{1}^yd\bar{y}\frac{3}{2}\sqrt{-\Gamma_j^2(\bar{y})}\right)^{\frac{2}{3}}.
\end{equation}
The uniform WKB ansatz is then
\begin{equation}
\Phi_{\lambda_j}(y)=C_1f(y)\mathrm{Ai}\left[\zeta_j(y)\right]+C_2g(y)\mathrm{Bi}\left[\zeta_j(y)\right]
\end{equation}
where $\mathrm{Ai}$ and $\mathrm{Bi}$ are Airy functions, and $f$ and $g$ are slowly varying functions to be determined. The normalizable solutions to the differential equation are as follows:
\begin{equation}
    \Phi_{\lambda_j}(y)=C_j\frac{\mathrm{Ai}[\zeta_j(y)]}{\sqrt{\zeta_j'(y)}},
\end{equation}
which represents an exponentially decaying function for $|y|>1$ and an oscillatory function for $|y|<1$. These solutions are WKB-like solutions to the differential equation in the entire range of $y$ only if the Bohr-Sommerfeld condition holds:
\begin{equation}
    \int_{-1}^{1}\Gamma_j(y)dy=\pi(\kappa+1/2),\qquad \kappa\in\mathbb{N}.
\end{equation}
Here, the integral is understood to avoid the small region $|y|<1/n_+$ where $\Gamma_j^2<0$. For large $n_+$
\begin{equation}
    \int_{-1}^{1}\Gamma_j(x)dx=2\alpha_j\int_{1/{n_+}}^{1}\sqrt{1/x^2-1}dx\sim 2\alpha_j\ln\left[2n_+\right]=\pi(\kappa+1/2).
\end{equation}
Then, solving for $\lambda_j$, one finds the following discrete spectrum:
\begin{equation}
\lambda_{j,\kappa}=\lambda_{j,0}\exp\left(-\frac{\kappa\pi}{\alpha_j}\right),\quad \kappa=0,1,...,
\end{equation}
where $\lambda_{j,0}$ is the ground level and depends on the cutoff for $\Gamma_j$ around $y=0$, given by Thiemann's trick. It is also sensitive to the approximation of the summation in $n$ by integrals in $y=n/n_+$. Notice that the Bohr-Sommerfeld condition implies parity on the eigenfunctions, which is shared by the index $\kappa$, that is, $\Phi_{\lambda_{j,\kappa}}$ is odd (even) if $\kappa$ is odd (even).

\section{Computation of \texorpdfstring{$|E^\varphi_j|$}{E phi j} matrix elements}\label{sec:Ephi_appendix}
In this section, we compute the matrix elements of $|E^\varphi_j|$ in the basis $\left\lbrace\left|\vec{\kappa}\right\rangle=\left|\kappa_1\right\rangle...\left|\kappa_1\right\rangle\right\rbrace$ introduced at the end of section \ref{sec:WKB} using the WKB approximation (\ref{eq:eigenfunctions}) for the eigenstates of the Hamiltonian constraint $\hat{C}_j$. We also confirmed the calculation by using a continuum limit exact solution. Because both $|E^\varphi_j|$ and $\hat{C}_j$ act non-trivially on the $\left|\vec{k},\vec{\mu}\right\rangle$ states only with respect to $\mu_j$, the matrix element of $|E^\varphi_j|$ is:
\begin{equation}\label{eq:_E_phi_matrix_element}
    \left\langle\vec{\kappa}\left||E_j^\varphi|\right|\vec{\kappa'}\right\rangle=\prod_{i\neq j}\delta_{\kappa_i,\kappa_i'}\left\langle\kappa_j\left||E_j^\varphi|\right|\kappa_j'\right\rangle=\prod_{i\neq j}\delta_{\kappa_i,\kappa'_i}\sum_{n\in\mathbb{Z}}\Phi_{\lambda_{j,{\kappa_j}}}^*(n)\left|2\rho n+\epsilon\right|\ell_{Pl}\Phi_{\lambda_{j,{\kappa_j'}}}(n),
\end{equation}
where $\mu_j=2\rho n+\epsilon$ and $\rho$ is the polimerization parameter.
To compute the summation, we used the fact that the main contribution comes from the values of $n$ around the turning points $n_\pm=\pm\alpha_j/\sqrt{-\lambda_j} $, which are large numbers for a macroscopic black hole. This allows us to use several approximations discussed in section \ref{sec:WKB} and in Appendix \ref{sec:WKB_appendix} . If both eigenfunctions have the same parity, then
\begin{equation}\label{eq:n}
\left\langle\kappa_j\left||E_j^\varphi|\right|\kappa_j'\right\rangle= 4 \rho\ell_{Pl}\sum_{n\geq0}n\Phi_{\lambda_{j,{\kappa_j}}}^*(n)\Phi_{\lambda_{j,{\kappa'_j}}}(n)\sim 4 \rho\ell_{Pl}(n_+)^2\int\limits_0^{+\infty}y\Phi_{\lambda_{j,{\kappa_j}}}^*(y)\Phi_{\lambda_{j,{\kappa'_j}}}(y)dy,
\end{equation}
where $y=n/n_+$ and the summation is approximated using integrals. If the parity differs, then
\begin{equation}\label{eq:epsilon}
    \left\langle\kappa_j\left||E_j^\varphi|\right|\kappa_j'\right\rangle\sim 4 \rho\ell_{Pl}\sum_{n\geq0}\epsilon\Phi_{\lambda_{j,{\kappa_j}}}^*(n)\Phi_{\lambda_{j,{\kappa'_j}}}(n)\sim4\rho\epsilon\ell_{Pl}n_+\int\limits_0^{+\infty}\Phi_{\lambda_{j,\kappa_j}}^*(y)\Phi_{\lambda_{j,{\kappa'_j}}}(y)dy.
\end{equation}
To start the computation, recall that the uniform WKB approximation to the eigenfunctions is described in Appendix \ref{sec:WKB_appendix}, and is given by
\begin{equation}
    \Phi_{\lambda_{j,{\kappa_j}}}(y)=C_j\frac{\mathrm{Ai}[\zeta_j(y)]}{\sqrt{\zeta_j'(y)}}.
\end{equation}
These functions peak at $y=1$, oscillate for $|y|<1$ and decay exponentially for $|y|>1$. Therefore, all integrals were divided into two pieces ($|y|\geq 1$ and $0\leq|y|<1$). From definition (\ref{eq:def_zeta}), we see that
\begin{align}
    \zeta_j(y)&=-\left(\alpha_j\int_{y}^1d\bar{y}\frac{3}{2}\sqrt{1/\bar{y}^2-1}\right)^{\frac{2}{3}}=-\left(\frac{3\alpha_j}{2}\right)^{\frac{2}{3}}\left[\mathrm{arcosh}\left(\frac{1}{y}\right)-\sqrt{1-y^2}\right]^{\frac{2}{3}},\quad 1>y>1/n_+\nonumber\\\zeta_j(y)&=\left(\alpha_j\int_{1}^yd\bar{y}\frac{3}{2}\sqrt{1-1/\bar{y}^2}\right)^{\frac{2}{3}}=\left(\frac{3\alpha_j}{2}\right)^{\frac{2}{3}}\left[\sqrt{y^2-1}-\arccos{\left(\frac{1}{y}\right)}\right]^{\frac{2}{3}},\quad y\geq1.
\end{align}
The small region $|y|<1/n_+$ is ignored in the limit $n_+>>1$ because it doesn't contribute significantly to the integrals.
\\
The first computation is the normalization constant
\begin{equation}
    |C_j|^{-2}=2n_+\int\limits_{0}^\infty{\frac{\mathrm{Ai}^2[\zeta_j(y)]}{\zeta_j'(y)}dy}.
\end{equation}
The leading contribution is from the region $y<1$, where the Airy function is oscillatory. With the change of variable $I=\mathrm{arcosh}\left(\frac{1}{y}\right)-\sqrt{1-y^2}$ we obtain:
\begin{equation}
    |C_j|^{-2}\sim\frac{2n_+}{\alpha_j}\int\limits_{0}^{+\infty}\left(\frac{3\alpha_j I}{2}\right)^{\frac{1}{3}}\mathrm{Ai}^2\left[-\left(\frac{3\alpha_j I}{2}\right)^{\frac{2}{3}}\right]\frac{y(I)^2}{1-y(I)^2}dI.
\end{equation}
Again, this integral can be split into the \textit{Airy region} ($3\alpha_jI/2<1$) where the argument of the Airy functions is small and the \textit{WKB region} ($3\alpha_jI/2>1$) where asymptotic expansions can be used. For large $\alpha_j$, the main contribution comes from the WKB region, where the Airy function is oscillatory, and the integral reduces to 
\begin{equation}
    |C_j|^{-2}\sim\frac{2n_+}{\alpha_j}\int\limits_{2/3\alpha_j}^{\infty}\frac{1}{\pi}\sin^2\left[\alpha_jI+\frac{\pi}{4}\right]\frac{y^2(I)}{1-y^2(I)}dI\sim\frac{n_+}{\pi\alpha_j}\int_0^1\frac{y}{\sqrt{1-y^2}}dy=\frac{n_+}{\pi\alpha_j}.
\end{equation}
In the middle step, the following identity is used to further isolate the main contribution to the integral, given by the non-oscillatory integrand:
\begin{equation*}
    \sin^2\left(\alpha_jI+\frac{\pi}{4}\right)=\frac{1}{2}+\sin(2\alpha_jI).
\end{equation*}
Next, consider the coincident case $\kappa_j=\kappa_j'=\kappa$ where only (\ref{eq:n}) must be calculated because (\ref{eq:epsilon}) is ruled out by parity. The integral is
\begin{equation}
     \left\langle\kappa\left||E_j^\varphi|\right|\kappa\right\rangle =4 \rho\ell_{Pl}(n_+)^2\int\limits_0^{+ \infty}y\left|\Phi_{\lambda_{j,\kappa}}(y)\right|^2dy=4 \rho\ell_{Pl}(n_+)^2|C_j|^2\int_0^{+\infty}x\frac{\mathrm{Ai}^2[\zeta_j(y)]}{\zeta_j'(y)}dy.
\end{equation}
Repeating the change of variable and split of the integrals performed previously, we obtain the following leading contribution:
\begin{equation}
     \left\langle\kappa\left||E_j^\varphi|\right|\kappa\right\rangle\sim4 \rho\ell_{Pl}(n_+)^2\left(\frac{n_+}{\pi\alpha_j}\right)^{-1}\frac{1}{2\pi\alpha_j}\int_0^1\frac{y^2}{\sqrt{1-y^2}}dy=2\rho\ell_{Pl}n_+\frac{\pi}{4}.
\end{equation}
The non-diagonal $\kappa\neq\kappa'$ terms are slightly more involved because the eigenfunctions are defined with respect to different eigenvalues. However, a technique similar to that used in the previous computation is available. Let us define $y=n/\sqrt{n_+n_+'}$, where $n_+=\alpha_j/\sqrt{-\lambda_{j,{\kappa}}}$ and $n_+'=\alpha_j/\sqrt{-\lambda_{j,{\kappa'}}}$. The corresponding eigenfunctions are as follows:
\begin{equation}
    \Phi_{\lambda_{j,\kappa}}(n)=C_j\frac{\mathrm{Ai}\left[\zeta\left(y\sqrt{n_+'/n_+}\right)\right]}{\zeta'\left(y\sqrt{n_+'/n_+}\right)},\qquad \Phi_{\lambda_{j,\kappa'}}(n)=\bar{C}_j\frac{\mathrm{Ai}\left[\zeta\left(y\sqrt{n_+/n_+'}\right)\right]}{\zeta'\left(y\sqrt{n_+/n_+'}\right)}.
\end{equation}
Using the exponential distribution of eigenvalues (\ref{eq:lambda_spectrum}) we define,
\begin{align}
    b&\equiv\sqrt{n_+/n_+'}=(\lambda_{j,\kappa'}/\lambda_{j,\kappa})^{1/4}=\exp([\kappa-\kappa']\pi/4\alpha_j),\\
    \bar{b}&\equiv\min(b,b^{-1}).
\end{align}
The starting point is the integral
\begin{equation}
    \left\langle\kappa\left||E_j^\varphi|\right|\kappa'\right\rangle\sim4 \rho\ell_{Pl}n_+n_+'C_j(\bar{C}_j)^*\int\limits_0^{+\infty}x\frac{\mathrm{Ai}\left[\zeta_j\left(yb^{-1}\right)\right]\mathrm{Ai}\left[\zeta_j\left(yb\right)\right]}{ \sqrt{\zeta_j'\left(yb^{-1}\right)\zeta_j'\left(yb\right)}}dy.
\end{equation}
With analogous changes of variables and separations of the integrals as before, we obtain the following for the leading contribution:
{\small\begin{align}
     &\left\langle\kappa\left||E_j^\varphi|\right|\kappa'\right\rangle\sim\frac{4 \rho\ell_{Pl}n_+n_+'C_j(\bar{C}_j)^*}{\pi\alpha_j}\int\limits_0^{\bar{b}}\sin\left[\alpha_j I(yb^{-1})+\frac{\pi}{4}\right]\sin\left[\alpha_j I(yb)+\frac{\pi}{4}\right]\frac{y^2 dy}{\left[1-y^2b^{-2}\right]^{\frac{1}{4}}\left[1-y^2b^2\right]^{\frac{1}{4}}}\\
    &=\frac{4 \rho\ell_{Pl}n_+n_+'C_j(\bar{C}_j)^*}{\pi\alpha_j}\int\limits_0^{\bar{b}}\frac{\cos\left(\alpha_j [I(yb^{-1})-I(yb)]\right)+\sin\left(\alpha_j [I(yb^{-1})+I(yb)]\right)}{2}\frac{y^2dy}{\left[1-y^2b^{-2}\right]^{\frac{1}{4}}\left[1-y^2b^2\right]^{\frac{1}{4}}}.
\end{align}}
The first cosine term within the integral is independent of $\alpha_j$ because $I(y)\sim\log(2/y)-1$ when $y$ is small; therefore, $I(yb^{-1})-I(yb)\sim\log(yb^{-1}/yb)\sim-[\kappa-\kappa']\pi/2\alpha_j$. For that reason, it provides the leading contribution to the integral in the $\alpha_j>>1$ limit. Using also that $b\sim1$:
\begin{equation}
     \left\langle\kappa\left||E_j^\varphi|\right|\kappa'\right\rangle\sim4 \rho\ell_{Pl}n_+n_+'\sqrt{\frac{\pi\alpha_j}{n_+}}\sqrt{\frac{\pi\alpha_j}{n'_+}}\frac{\cos\left([\kappa-\kappa']\pi/2\right)}{2\pi\alpha_j}\int\limits_{0}^{1}\frac{y^{2}}{\sqrt{1-y^2}}dy.
\end{equation}
The remaining integral is the same as we had for the expectation value of $|E_j^\varphi|$ so the matrix element is
\begin{equation}
\left\langle\kappa\left||E_j^\varphi|\right|\kappa'\right\rangle\sim2 \rho\ell_{Pl}\sqrt{n_+n_+'}\cos\left([\kappa-\kappa']\pi/2\right)\frac{\pi}{4}
\end{equation}
Returning to (\ref{eq:_E_phi_matrix_element}), we have found that the matrix element is
\begin{equation}\label{eq:E_matrix_element}\left\langle\vec{\kappa}\left||E_j^\varphi|\right|\vec{\kappa'}\right\rangle=\frac{\pi}{4}\frac{|x_j|\cos([\kappa_j-\kappa'_j]\pi/2)}{\left(1-\frac{R_S-l_{j,{\kappa_j}}}{|x_j|}\right)^{\frac{1}{4}}\left(1-\frac{R_S-l_{j,{\kappa_j'}}}{|x_j|}\right)^{\frac{1}{4}}}
\end{equation}
where we have used $\alpha_j\sim\frac{|x_j|}{\ell_{Pl}}>>1$. Finally, for eigenfunctions of different parities, the required computation is 
    \begin{equation}
    \left\langle\kappa\left||E_j^\varphi|\right|\kappa'\right\rangle=4\rho\epsilon\ell_{Pl}n_+\int\limits_0^{\infty}\Phi_{\lambda_{j,\kappa}}^*(n)\Phi_{\lambda_{j,\kappa'}}(n)dn,
    \end{equation}
    which is similar to the previous integral, but with one power of $n$ missing. This implies a power of $n_+$ less in the final result and a reduction of the leading terms by a power of $1/\alpha_j$, making all its contributions sub-leading. The previous expression already considers this case to the leading order because $\cos\left([\kappa-\kappa']\pi/2\right)$ vanishes whenever $\kappa$ and $\kappa'$ are of different parity.
\subsection{Comparison with the exact solution in the continuum limit}
The WKB method gives a good approximation for the spectrum and qualitative features of the eigenfunctions, but it can introduce important numerical errors in the limit $\alpha_j>>1$, because the approximation breaks when the amplitude and phase of the eigenfunctions change rapidly in $n$. However, under the same approximations used to do computations in the WKB formalism, we have exact solutions to the eigenvalue equation provided  $y=n/n_+>>1/n_+$. Writing $\Phi_{\lambda_j}(y)=\sqrt{y}\varphi(\alpha_jy)$, the differential equation (\ref{eq:dif_equation}) becomes
\begin{equation}
    z^2\frac{d^2\varphi}{dz^2}(z)+z\frac{d\varphi}{dz}(z)-\left(z^2+\frac{1}{4}-\alpha_j^2\right)\varphi(z)=0,
\end{equation}
where $z=\alpha_jy$. This is the modified Bessel equation of order $\nu=i\sqrt{\alpha_j^2-1/4}$. Therefore, the normalizable eigenfunctions are:
\begin{equation}
    \Phi_{\lambda_j}(n)=A^{\pm}_j\sqrt{n/n_+}K_\nu(\alpha_jn/n_+),\quad |n|>1,
\end{equation}
where $A^\pm_j$ correspond to the normalization constants for the regions $n>+1$ and $n<-1$ respectively. The separation of the two regions corresponds to the divergence at $n=0$ of the asymptotic potential. We need to split them into even and odd combinations to match them one to one with the solutions of the original bounded potential.
For any parity and ignoring the contribution of the $x<1/n_+$ region, we have (see Section 6.576, formula 4 in \cite{gradshteyn2007})
\begin{align}
|A_j|^{-2}&=\sum_n|\Phi_{\lambda_j}(n)|^2=2n_+\int_0^{+\infty}dyy|K_\nu(\alpha_jy)|^2=\\
&=\frac{n_+}{\alpha_j^{2}}\Gamma(1+\nu)\Gamma(1-\nu)F(1+\nu,1,2,0)=\frac{n_+}{\alpha_j^{2}}\frac{\pi|\nu|}{\sinh(\pi|\nu|)}.
\end{align}
Here, $F$ is the Gaussian hypergeometric function, $\Gamma$ is the Euler gamma function, and the identity $|\Gamma(1+ia)|^2=\pi a/\sinh(\pi a)$ is used in the last step.\\
By using the same identity for the integral of the modified Bessel functions, we can compute the matrix element of $E^\varphi_j$. As in the previous subsection, the computation is reduced to 
\begin{align}
    &\left\langle\Phi_{\lambda_j}\left||E_j^\varphi|\right|\Phi_{\lambda'_j}\right\rangle=4 \rho\ell_{Pl}(n_+)^2\int\limits_{0}^\infty y\Phi_{\lambda_j}^*(y)\Phi_{\lambda'_j}(y)dy=\\
    &=4 \rho\ell_{Pl}(n_+)^2A_j^*A_j'\int\limits_0^\infty y^2\left(\frac{\lambda_j'}{\lambda_j}\right)^{1/4}K_\nu(\alpha_jy)K_\nu\left(\alpha_jy\sqrt{\frac{\lambda'_j}{\lambda_j}}\right)dy=\\
    &=4 \rho\ell_{Pl}\frac{(n_+)^2}{\alpha_j^3}\frac{A_j^*A_j'}{\Gamma(3)}\left(\frac{\lambda_j'}{\lambda_j}\right)^{\nu/2+1/4}\Gamma\left(\frac{3}{2}+\nu\right)\left[\Gamma\left(\frac{3}{2}\right)\right]^2\Gamma\left(\frac{3}{2}-\nu\right)F\left(\frac{3}{2}+\nu,\frac{3}{2},3,1-\frac{\lambda_j'}{\lambda_j}\right)=\\
    &=4 \rho\ell_{Pl}\frac{(n_+)^2}{\alpha_j^3}\frac{A_j^*A_j'}{2}\left(\frac{\lambda_j'}{\lambda_j}\right)^{\nu/2+1/4}\frac{\pi}{4}\frac{\pi(1/4+|\nu|^2)}{\cosh(\pi|\nu|)}F\left(\frac{3}{2}+\nu,\frac{3}{2},3,1-\frac{\lambda_j'}{\lambda_j}\right)=\\
    &=2 \rho\ell_{Pl}\sqrt{n_+n_+'}\frac{\pi}{4}\tanh(\pi|\nu|)\frac{\alpha_j}{|\nu|}\left(\frac{\lambda_j'}{\lambda_j}\right)^{\nu/2+3/4}F\left(\frac{3}{2}+\nu,\frac{3}{2},3,1-\frac{\lambda_j'}{\lambda_j}\right).
\end{align}
Here, the identity $|\Gamma(\frac{3}{2}+ia)|^2=\frac{\pi}{\cosh(\pi a)}(a^2+1/4)$ is used, and the expression is written to highlight the symmetry under the exchange $\lambda_j\leftrightarrow \lambda_j'$.
In the coincident case ($\lambda_j=\lambda_j'$) it reduces to
\begin{equation}
    \left\langle\Phi_{\lambda_j}\left||E_j^\varphi|\right|\Phi_{\lambda_j}\right\rangle=2 \rho\ell_{Pl}n_+\frac{\pi}{4}\tanh(\pi|\nu|)\frac{\alpha_j}{|\nu|}\sim 2 \rho\ell_{Pl}n_+\frac{\pi}{4}\quad (\alpha_j>>1)
\end{equation}
When $\lambda_j'\neq\lambda_j$, the limit $\alpha_j>>1$ in the hipergeometric function is not uniform with respect to $\lambda_j'/\lambda_j$, but we can simplify the prefactor to obtain
\begin{equation}  \left\langle\Phi_{\lambda_j}\left||E_j^\varphi|\right|\Phi_{\lambda'_j}\right\rangle=\frac{|x_j|}{\left(1-\frac{R_S-l_j}{|x_j|}\right)^{\frac{1}{4}}\left(1-\frac{R_S-l_j'}{|x_j|}\right)^{\frac{1}{4}}}\frac{\pi}{4}J(\alpha_j,\lambda_j'/\lambda_j)
\end{equation}
where $\lambda_j$ and $l_j$ are related using (\ref{eq:lambda_def}). The function $J$ is defined as
\begin{equation}
    J(a,b)=b^{3/4+ia/2}F\left(\frac{3}{2}+ia,\frac{3}{2},3,1-b\right).
\end{equation}
Two of its relevant properties are $J(a,1)=1$ and $J(a,b)=J(a,1/b)$. To go further, we need to introduce the relation (\ref{eq:lambda_spectrum}), that is, $\lambda_j'/\lambda_j=\lambda_{j,\kappa'}/\lambda_{j,\kappa}\sim\exp([\kappa-\kappa']\pi/\alpha_j)$ which does not appear directly in the continuum limit. The case of interest is $[\kappa-\kappa']\alpha_j<<1$, where the eigenvalues are close together. For that regime
    \begin{equation}
        J(\alpha_j,\exp([\kappa-\kappa']\alpha_j))\sim\cos([\kappa-\kappa']\pi/2)
    \end{equation}
and we finally obtain
\begin{equation}  \left\langle\vec{\kappa}\right||E_j^\varphi|\left|\vec{\kappa}'\right\rangle\sim\frac{\pi}{4}\frac{|x_j|\cos([\kappa-\kappa']\pi/2)}{\left(1-\frac{R_S-l_{j,\kappa}}{|x_j|}\right)^{\frac{1}{4}}\left(1-\frac{R_S-l_{j,\kappa'}}{|x_j|}\right)^{\frac{1}{4}}},
\end{equation}
which is the same result we obtained with the WKB approximation.
%
%

\end{document}